\documentclass{article}
\usepackage{axodraw}
\newcommand\GP{G_{1} \! \! \left( \! z,\frac{m_{h}^{2}}{Q^{2}}\right)}
\newcommand\GSC{G_{2} \! \! \left( \! z,\frac{m_{h}^{2}}{Q^{2}}\right)}
\newcommand\rrad{\sqrt{1+\frac{4z}{z-1} \frac{m_{h}^{2}}{Q^{2}}}}
\newcommand\lloga{\ln \left( \frac{1+\rrad}{1-\rrad} \right)}
\begin{document}
\large
\title{Heavy flavours production in deeply inelastic scattering and gluon
density in the proton}
\author{\Large P. Balbi, A. Giovannini \\
\Large Theoretical Physics Department, University of Torino}
\maketitle
Report-no: DFTT 40/02
\begin{center}
{\bf Abstract}
\end{center}
Heavy flavours production in e-p DIS is studied at intermediate values of the
transferred four-momentum square, under the assumption of boson-gluon-fusion
mechanism dominance (no intrinsic heavy flavours contributions). \\
In this framework different expressions for the splitting functions in the
gluon density evolution equation, with respect to the standard (DGLAP) ones,
are explicitly derived. \\
\newpage
Lepton-proton deeply inelastic scattering can be studied perturbatively in the
single-photon-exchange approximation neglecting electroweak corrections (\(
Z_{0} \) exchange) when \( \Lambda^{2}_{QCD} \ll Q^{2} \ll M_{Z}^{2} \) (\(
q^{2} = -Q^{2} \) is the square of the transferred four momentum).\\
The following expression for the differential cross section, with respect to
\( Q^{2} \) and the Bjorken scaling variable \( x \), can be obtained by
Lorentz-invariance and current conservation arguments:
\begin{equation}
\frac{d^{2} \sigma}{dx dQ^{2}} = \frac{2 \pi \alpha^{2}}{x Q^{4}} \left[
2 x y^{2} F_{1} \left( x , Q^{2} \right) + 2 \left( 1 - y \right) F_{2} \left(
x , Q^{2} \right) \right]
\end{equation}
\( \alpha \) is the e.m fine structure constant \( \left( \alpha \simeq
\frac{1}{137} \right) \) and \( y \) is the fraction of energy lost by the
lepton, during the collision, in the proton rest frame. \\
The proton electromagnetic structure functions \( F_{1,2} \left( x,Q^{2} 
\right) \) satisfy the following factorization formulas \cite{fact1}
\cite{fact2} \cite{fact3}
\begin{eqnarray}
F_{1} \left(x , Q^{2} \right) = \sum_{i} \int_{x}^{1} \frac{d \xi}{\xi}
C^{(i)}_{1} \left( \frac{x}{\xi} , \frac{Q^{2}}{\mu^{2}} , \alpha_{s} \left(
\mu^{2} \right) \right) \rho_{i} \left( \xi , \epsilon , \mu^{2} \right) \\
F_{2} \left(x , Q^{2} \right) = \sum_{i} \int_{x}^{1} d \xi
C^{(i)}_{2} \left( \frac{x}{\xi} , \frac{Q^{2}}{\mu^{2}} , \alpha_{s} \left(
\mu^{2} \right) \right) \rho_{i} \left( \xi , \epsilon , \mu^{2} \right)
\end{eqnarray}
where \( \mu \) is the renormalization scale and \( \rho_{i} \) is the density
of the partonic species \( i \) in the proton with respect to the fraction of
longitudinal momentum carried by the parton itself \( \left( \xi \right) \),
while \( \epsilon \) stands for the infrared cutoff used to regularize
collinear divergences. \\
The coefficients \( C^{(i)}_{1,2} \left( \frac{x}{\xi} , \frac{Q^{2}}{\mu^{2}}
, \alpha_{s} \left( \mu^{2} \right) \right) \) take into account the
short-distance effects, they are infrared safe because of a Bloch-Nordsieck
type compensation of soft divergences \cite{fact1} \cite{blonor} and can be
computed in perturbative QCD. \\
The long-distance dynamics is enterely factorized in the parton densities
\( \rho_{i} \left( \xi , \epsilon , \mu^{2} \right) \). \\
The scale independence condition for the structure function \( F_{2} \) 
\[ \frac{\partial F_{2} \left(x , Q^{2} \right)}{\partial \log \mu^{2}} = 0 \]
leads to the renormalization group equations for the quarks densities, id est
\[
\frac{\partial \rho_{f} \left( x , \epsilon , \mu^{2} \right)}{\partial \log
\mu^{2}} = \frac{\alpha_{s} \left( \mu^{2} \right)}{2 \pi} \int_{x}^{1}
\frac{d \xi}{\xi} \left[ P_{ff} \left( \frac{x}{\xi} \right) \rho_{f} \left( 
\xi , \epsilon , \mu^{2} \right) + \right. \]
\begin{equation}
\left. P_{fg} \left( \frac{x}{\xi} \right) \rho_{g}
\left( \xi , \epsilon , \mu^{2} \right) \right]
\end{equation}
where \( \alpha_{s} \left( \mu^{2} \right) \) is the strong running coupling
constant. \\
The explicit expressions of the splitting functions \( P_{ff} \left( z \right)
\) and \( P_{fg} \left( z \right) \) are found to be \cite{dglap1}
\cite{dglap2}
\begin{equation}
P_{ff} \left( z \right) = C_{2} \left( F \right) \left( \frac{1+z^{2}}{1-z}
\right)_{+} \ , \ P_{fg} \left( z \right) = T \left( F \right) \left[ z^{2} +
\left( 1 - z \right)^{2} \right]
\end{equation}
where in general \( \left[ F \left( z \right) \right]_{+} \) is a distribution
defined as
\[ \int_{x}^{1} dz \left[ F \left( z \right) \right]_{+} \phi \left( z \right)
= \int_{x}^{1} dz F \left( z \right) \left[ \phi \left( z \right) - \phi \left(
 1 \right) \right] \]
and
\[ C_{2} \left( F \right) = \frac{N^{2} - 1}{2 N} = \frac{4}{3} \ , \ T \left(
F \right) = \frac{1}{2} \]
Charm production in \( e^{+} p \) DIS has been extensively studied by ZEUS and
H1 collaborations at HERA electron-proton collider \cite{dstar1} \cite{dstar2}
\cite{dstar3} \cite{dstar4}. \\
In particular the charm contribution to the electromagnetic structure functions
of the proton is estimated from the measurements of the cross sections for
inclusive \( D^{* \pm} \) production, given the value for the hadronization
fraction of a quark charm in \( D^{* +} \) by OPAL Collab. (\( e^{+} \ e^{-} \)
annihilation \cite{hadfra}):
\[ f \left( c \rightarrow D^{* +} \right)  = 0.222 \pm 0.014 \pm 0.014 \]
Experimental data can be explained in perturbative QCD by assuming that the
proton wave function does not possess any heavy flavour content; under this
hypothesis heavy quarks production in e-p DIS is thought to be governed
by boson-gluon-fusion (BGF) pair production according to the partonic
subprocess (figure 1)
\[ \gamma^{*} \ g \rightarrow q_{h} \ \bar{q}_{h} \ X \]
That kind of experiments provides important informations on the gluonic
structure of the proton; actually the most precise measurements of the gluon
density in the proton available nowadays are coming from HERA experiments
\cite{dstar4} \cite{glu}. \\
The differential cross section for inclusive production of a pair of heavy
quarks in virtual photon-proton scattering
\[ \gamma^{*} \left( q \right) \ p \left( P \right) \rightarrow q_{h}
\bar{q}_{h} \left( M^{2} \right) \ X \] 
in pair invariant mass kinematics, enjoys the following factorization theorem
\[ \frac{d \sigma^{(h \bar{h})} \left( x, M^{2} , Q^{2} , m_{h}^{2} \right)
}{d M^{2}} = \frac{x^{2}}{Q^{4}} \sum_{i} \int_{x}^{1} \frac{d \xi}{\xi} 
\rho_{i} \left( \xi , \epsilon , \mu^{2} \right) \] 
\begin{equation}
\hat{\omega}_{i} \left( \frac{x}{\xi} , 
\frac{Q^{2}}{\mu^{2}} , \frac{M^{2}}{\mu^{2}} , \frac{m_{h}^{2}}{\mu^{2}} ,
\alpha_{s} \left( \mu^{2} \right) \right)
\end{equation}
where \( m_{h} \) represents the heavy flavour mass, \( M \) is the invariant
mass of the pair while the sum is understood to be over all massless partonic
species (gluon and light (anti)quarks).\\
The coefficient functions \( \hat{\omega}_{i} \! \left( \frac{x}{\xi} , 
\frac{Q^{2}}{\mu^{2}} , \frac{M^{2}}{\mu^{2}} , \frac{m_{h}^{2}}{\mu^{2}} ,
\alpha_{s} \left( \mu^{2} \right) \right) \! \) include, as usual, the
short-distance dynamics; they turn out to be infrared safe (compensation of
soft divergences) and can be computed in perturbative QCD. \\
The collinear divergences are regularized and factorized in the parton
densities \( \rho_{i} \left( \xi , \epsilon , \mu^{2} \right) \). \\
\begin{center} \begin{picture}(-200,-25)(200,100)
\Line(100,50)(122,61)
\Gluon(20,90)(78,61){2.5}{6}
\Line(78,61)(100,50)
\ArrowLine(122,61)(180,90)
\Photon(20,10)(100,50){2.5}{10}
\ArrowLine(190,65)(120,55)
\CCirc(100,50){30}{}{}
\Text(20,0)[uc]{\( \gamma^{*} \)}
\Text(20,100)[dc]{\( g \)}
\Text(190,95)[dc]{\( q_{h} \)}
\Text(200.5,65.15)[uc]{\( \bar{q}_{h} \)}
\Line(129.5,45)(175,30)
\Line(128,39)(165,15)
\Line(175,30)(180,37.5)
\Line(165,15)(160,7.5)
\Line(160,7.5)(184.5,15)
\Line(180,37.5)(184.5,15)
\Text(192,14)[dc]{\( X \)}
\Text(106,-25)[cc]{Figure 1: BGF mechanism}
\end{picture} \end{center}
\vspace{1.9 in}
All physical quantities turn out to be scale independent, then from the
condition
\[ \frac{\partial}{\partial \log \mu^{2}} \frac{d \sigma^{\left( h \bar{h}
\right)} \left( x , M^{2} , Q^{2} , m_{h}^{2} \right)}{d M^{2}} = 0 \]
one may obtain the renormalization group equation for the density of gluons;
at leading order it gives
\[ \frac{\partial \rho_{g} \left( x , \epsilon , \mu^{2} \right)}{\partial 
\log \mu^{2}} = \frac{\alpha_{s} \left( \mu^{2} \right)}{2 \pi} \int_{x}^{1} 
\frac{d \xi}{\xi} \left[ \sum_{f} P_{gf} \left( \frac{x}{\xi} \right) \rho_{f} 
\left( \xi , \epsilon , \mu^{2} \right) + \right. \]
\begin{equation}
\left. P_{gg} \left( \frac{x}{\xi} \right) \rho_{g} \left( \xi , \epsilon , 
\mu^{2} \right) \right]
\end{equation} 
The Born term of the lepton-proton cross sections for heavy flavours production
in deeply inelastic scattering coincides with the lowest order contribution of
the transition
\[ e \ g \rightarrow e \ q_{h} \ \bar{q}_{h} \]
It comes from the partonic diagrams in figure 2:
\begin{center} \begin{picture}(350,180)(10,0)
\Gluon(20,130)(70,130){3}{7.5}
\Gluon(200,130)(250,130){3}{7.5}
\ArrowLine(160,166)(70,130)
\ArrowLine(70,130)(90,122)
\ArrowLine(20,30)(90,58)
\ArrowLine(90,58)(160,30)
\ArrowLine(250,130)(340,94)
\ArrowLine(270,138)(250,130)
\ArrowLine(200,30)(270,58)
\ArrowLine(270,58)(340,30)
\ArrowLine(340,170)(270,138)
\ArrowLine(90,122)(160,94)
\Photon(90,122)(90,58){3}{7.5}
\Photon(270,58)(270,118){3}{7.5}
\Photon(270,126)(270,138){3}{1.5}
\CArc(270,122)(4,-90,90)
\Vertex(70,130){1.6}
\Vertex(92,121.5){1.6}
\Vertex(90,58){1.6}
\Vertex(268,137.5){1.6}
\Vertex(270,58){1.6}
\Vertex(250,130){1.6}
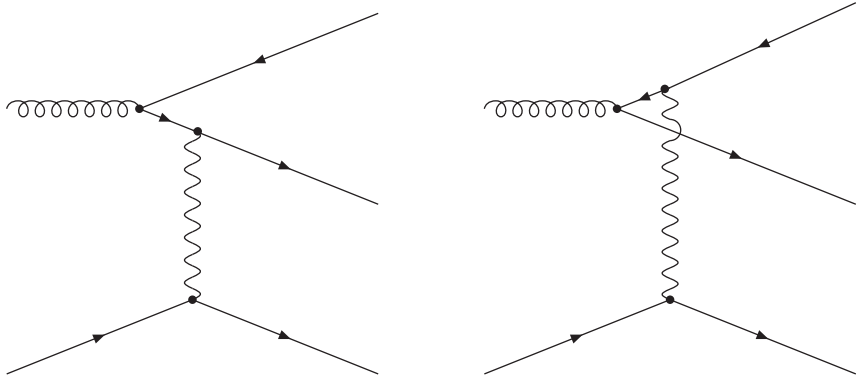
\Text(180,5)[cc]{Figure 2: Born terms for \( e \ g \rightarrow e \ q_{h} \
\bar{q}_{h} \)}

\end{picture} \end{center}
\vspace{0.15 in}
The calculations are straightforward and have already been performed in
references \cite{born} \cite{hf}. \\
This contribution to the double-differential cross section \( \frac{d^{2}
\sigma}{dx dQ^{2}} \) can be written in the same way as in eq.(1):
\begin{equation}
\frac{d^{2} \sigma^{\left( h \bar{h} \right)}_{B}}{dx dQ^{2}} = \frac{2 \pi 
\alpha^{2}}{x Q^{4}} \left[ 2 x y^{2} F_{1}^{h \bar{h}} \left(x , Q^{2} , m_{h}
\right) + 2 \left( 1 - y \right) F_{2}^{h \bar{h}} \left(x , Q^{2} , m_{h}
\right) \right] 
\end{equation}
where  \( F_{1,2}^{h \bar{h}} \left( x , Q^{2} , m_{h}
\right) \) are the lowest order contributions of the heavy flavour \( h \) to
the e.m structure functions of the proton. \\
They can be expressed as follows
\[ F_{1}^{h \bar{h}} \! \left(x,Q^{2},m_{h} \right) \! = \! 
\frac{\alpha_{s}Q_{h}^{2}}{4 \pi} \!
\int_{z_{min}}^{z_{max}} \! \frac{dz}{z} \! \rho_{g} \left( \frac{x}{z},\mu^{2}
\right) \! \! \left[ \GP \! - \! \GSC \! \right] \]
\[ F_{2}^{h \bar{h}} \! \left(x,Q^{2},m_{h} \right) = \] 
\[ \frac{\alpha_{s}Q_{h}^{2}x}{2 \pi} \!
\int_{z_{min}}^{z_{max}} \! \frac{dz}{z} \! \rho_{g} \left( \frac{x}{z},\mu^{2}
\right) \! \left[ \GP \! - \! 3 \GSC \! \right] \]
with \( z_{min} = x \), \( z_{max} = \frac{Q^{2}}{Q^{2} + 4 m_{h}^{2}} \) and
\[ \GP \! = \!
- \! \left[ z^{2} + \left(1 - z \right) \left(1 - z +
\frac{4zm_{h}^{2}}{Q^{2}} \right) \right] \! \rrad  \]
\[ + \left[z^{2} + \left(1 - z \right)^{2} +
\frac{4zm_{h}^{2}}{Q^{2}} \left(1 - \frac{2zm_{h}^{2}}{Q^{2}}
\right) \right] \lloga \]
\[ \GSC = \]
\[ 2z^{2} \left\{ \frac{2m_{h}^{2}}{Q^{2}} \lloga + \frac{z - 1}{z}
\rrad \right\} \]
while \( Q_{h} \) is the heavy quark electric charge (in e units). \\
The aim of this paper is to derive the explicit form of the splitting functions
 \( P_{gf} \left( z \right) \) and \( P_{gg} \left( z \right) \) which appear
in the evolution equation for the gluon density, directly from the study of
heavy flavours production in e-p DIS. \\
Then these results are compared with the corresponding quantities obtained by
Altarelli and Parisi in a different framework \cite{dglap1} \cite{dglap2}. \\
For this purpose we start by considering the next-to-leading order (NLO), quark
initiated corrections to the Born term, namely the lowest order contributions
to 
\[ e \ q_{f} \rightarrow e \ q_{f} \ q_{h} \ \bar{q}_{h} \]
where \( q_{f} \) is here any light quark. They are given by the diagrams shown
in figures 3 and 4 (all the calculations in the present paper are made in
Feynman Gauge). \\ 
The corresponding contribution to the differential electron-proton \(
\frac{d^{2} \sigma}{dx dQ^{2}} \) cross section is expressed, in the framework
of free parton model, as a convolution over the free (light)quark density
\[ \frac{d^{2} \sigma^{\left( h \bar{h} \right)}_{f}}{dx dQ^{2}} = 
\int_{x}^{1} \frac{d \xi}{\xi} \rho_{f} \left( \xi \right) \frac{d^{2} 
\hat{\sigma}^{\left( h \bar{h} \right)}_{f}}{dz dQ^{2}} 
\]
where \( z = \frac{x}{\xi} \) represents the ``partonic'' Bjorken
variable and \( \frac{d^{2} \hat{\sigma}_{f}^{\left( h \bar{h} \right)}}{dz
dQ^{2}} \) is the corresponding lepton-parton cross section. \\
\vspace{1.12 in}
\begin{center}  \begin{picture}(400,30)(20,-105)
\ArrowLine(40,15)(70,0)
\ArrowLine(90,-10)(130,-30)
\ArrowLine(210,10)(250,-10)
\ArrowLine(270,-20)(310,-40)
\ArrowLine(250,-10)(270,-20)
\ArrowLine(180,10)(140,-10)
\ArrowLine(140,-10)(180,-30)
\ArrowLine(70,0)(90,-10)
\ArrowLine(340,10)(300,-10)
\ArrowLine(300,-10)(340,-30)
\ArrowLine(30,-90)(70,-70)
\ArrowLine(70,-70)(110,-90)
\ArrowLine(230,-90)(270,-70)
\ArrowLine(270,-70)(310,-90)
\Photon(70,0)(70,-70){2.5}{8}
\Photon(270,-20)(270,-70){2.5}{6}
\Gluon(90,-10)(140,-10){2.5}{6}
\Gluon(250,-10)(300,-10){2.5}{6}
\Vertex(70,0){1.6}
\Vertex(70,-70){1.6}
\Vertex(270,-20){1.6}
\Vertex(270,-70){1.6}
\Vertex(90,-10){1.6}
\Vertex(140,-10){1.6}
\Vertex(250,-10){1.6}
\Vertex(300,-10){1.6}
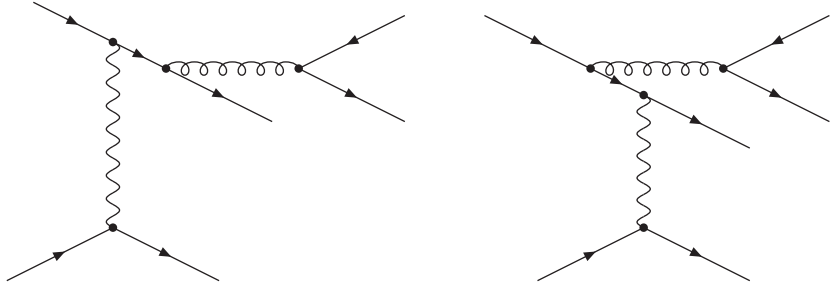
\Text(186,-120)[cc]{Figure 3: NLO quark-initiated corrections; the e.m}
\Text(186,-135)[cc]{interaction involves the light quark}
\end{picture}  \end{center}
\vspace{0.28 in}
\begin{center}  \begin{picture}(400,140)(25,45)
\ArrowLine(30,160)(60,145)
\ArrowLine(60,145)(90,130)
\ArrowLine(170,175)(110,145)
\ArrowLine(110,145)(140,130)
\ArrowLine(140,130)(170,115)
\ArrowLine(100,60)(140,80)
\ArrowLine(140,80)(180,60)
\ArrowLine(210,160)(240,145)
\ArrowLine(240,145)(270,130)
\ArrowLine(320,160)(290,145)
\ArrowLine(350,175)(320,160)
\ArrowLine(290,145)(350,115)
\ArrowLine(280,60)(320,80)
\ArrowLine(320,80)(360,60)
\Photon(140,80)(140,130){2.0}{5.5}
\Photon(320,80)(320,125){2.0}{4.5}
\Photon(320,135)(320,160){2.0}{2.5}
\Gluon(60,145)(110,145){2.0}{5.5}
\Gluon(240,145)(290,145){2.0}{5.5}
\CArc(317,130)(6,-60,60)
\Vertex(320,80){1.6}
\Vertex(320,160){1.6}
\Vertex(140,80){1.6}
\Vertex(140,130){1.6}
\Vertex(60,145){1.6}
\Vertex(110,145){1.6}
\Vertex(240,145){1.6}
\Vertex(290,145){1.6}
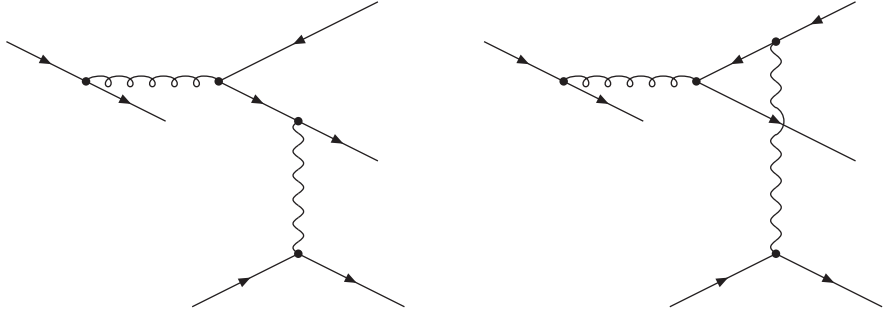
\Text(197,32)[cc]{Figure 4: NLO quark-initiated corrections with the}
\Text(197,16)[cc]{virtual photon probing a heavy flavour}
\end{picture}  \end{center}
\newpage
One can easily realize that the collinear divergent part of \( \frac{d^{2}
\sigma^{\left( h \bar{h} \right)}_{f}}{dx dQ^{2}} \) comes from the last two
diagrams in figure 4; it can be written in the same form as the Born
contribution
\begin{equation}
\frac{d^{2} \sigma^{\left( h \bar{h} \right)}_{B}}{dx dQ^{2}
} = \int_{0}^{1} \frac{d \xi}{\xi} \rho_{g} \left( \xi \right) \frac{d^{2} 
\hat{\sigma}^{\left( h \bar{h} \right)}_{B}}{dz dQ^{2}}
\end{equation}
simply by subsituting the ``free'' gluon density \( \rho_{g} \left( \xi
\right) \) with its next-to-leading order correction \cite{phd}
\[ \delta \rho^{(1)}_{g} \left( \xi , \frac{Q^{2}}{Q^{2}_{0}} \right) =
\frac{\alpha_{s}}{2 \pi} \log \left(\frac{Q^{2}}{Q^{2}_{0}} \right) \int_{\xi}^
{1} \frac{d \lambda}{\lambda} P_{gf} \left( \frac{\xi}{\lambda} \right)
\rho_{f} \left( \lambda \right)  \]
where \( \frac{Q}{Q_{0}} \) is the infrared cutoff used to regularize the
light quark-gluon collinear singularities (we choose \( Q \) as
factorization/renormali-
zation scale). \\
Therefore we can regularize these NLO collinear divergences and reabsorb them
in the Born term through the following redefinition of the gluon density in
eq.(9)
\begin{equation}
\rho_{g} \left( \xi \right) \Rightarrow \rho_{g} \left( \xi \right) +
\delta \rho_{g}^{(1)} \left( \xi , \frac{Q^{2}}{Q_{0}^{2}} \right)
\end{equation}
The splitting function \( P_{gf} \left( z \right) \) is given by
\begin{equation}
P_{gf} \left( z \right) = C_{2} \left( F \right) z = \frac{4}{3} z
\end{equation}
It should be noticed that this expression vanishes linearly in the limit
\( z \to 0 \) and that it is different from the following formula obtained in
references \cite{dglap1} \cite{dglap2}
\[ P_{gf} \left( z \right) = C_{2} \left( F \right) \frac{1 + \left( 1 - z
\right)^{2}}{z} \]
To compute now the explicit expression for the gluon-gluon splitting
function \( P_{gg} \left( z \right) \),  the next-to-leading order, gluon
initiated corrections, due to real gluon emission, have been derived in the
collinear limit. \\
Essentially the following partonic process has to be studied (at lowest order):
\[ e \ g \rightarrow e \ g \ q_{h} \ \bar{q}_{h} \]
There are in total 8 diagrams to consider, they are shown in figures 5 and 6: 
\newpage
\ \ \ \\
\begin{center} 
\begin{picture}(360,290)(20,0)
\ArrowLine(80,250)(120,270)
\ArrowLine(120,270)(160,290)
\ArrowLine(110,235)(80,250)
\ArrowLine(160,210)(110,235)
\ArrowLine(60,160)(110,185)
\ArrowLine(110,185)(160,160)
\Gluon(30,250)(80,250){2.5}{6}
\Gluon(120,270)(160,270){-2.0}{5}
\Photon(110,185)(110,235){2.0}{6}
\Vertex(110,185){1.6}
\Vertex(110,235){1.6}
\Vertex(80,250){1.6}
\Vertex(120,270){1.6}
\ArrowLine(260,250)(340,290)
\ArrowLine(300,230)(280,240)
\ArrowLine(340,210)(300,230)
\ArrowLine(280,240)(260,250)
\ArrowLine(260,160)(300,180)
\ArrowLine(300,180)(340,160)
\Gluon(210,250)(260,250){2.5}{6}
\Gluon(280,240)(330,240){2.5}{6}
\Photon(300,180)(300,230){2.0}{6}
\Vertex(300,180){1.6}
\Vertex(300,230){1.6}
\Vertex(260,250){1.6}
\Vertex(280,240){1.6}
\ArrowLine(80,90)(160,130)
\ArrowLine(110,75)(80,90)
\ArrowLine(135,62.5)(110,75)
\ArrowLine(160,50)(135,62.5)
\ArrowLine(60,0)(110,25)
\ArrowLine(110,25)(160,0)
\Gluon(30,90)(80,90){2.5}{6}
\Gluon(135,62.5)(165,62.5){2.0}{4}
\Photon(110,25)(110,75){2.0}{6}
\Vertex(110,25){1.6}
\Vertex(110,75){1.6}
\Vertex(80,90){1.6}
\Vertex(135,62.5){1.6}
\ArrowLine(260,90)(290,105)
\ArrowLine(290,105)(315,117.5)
\ArrowLine(315,117.5)(340,130)
\ArrowLine(340,50)(260,90)
\ArrowLine(240,0)(290,25)
\ArrowLine(290,25)(340,0)
\Gluon(210,90)(260,90){2.5}{6}
\Gluon(315,117.5)(345,117.5){-2.0}{4}
\Photon(290,80)(290,105){2.0}{3.5}
\Photon(290,25)(290,70){2.0}{5.5}
\CArc(290,75)(5,-90,90)
\Vertex(290,25){1.6}
\Vertex(290,105){1.6}
\Vertex(315,117.5){1.6}
\Vertex(260,90){1.6}
\ArrowLine(80,-70)(100,-60)
\ArrowLine(100,-60)(120,-50)
\ArrowLine(160,-110)(80,-70)
\ArrowLine(120,-50)(160,-30)
\ArrowLine(80,-150)(120,-130)
\ArrowLine(120,-130)(160,-150)
\Gluon(30,-70)(80,-70){2.5}{6}
\Gluon(100,-60)(115,-60){-2.5}{2.0}
\Gluon(125,-60)(160,-60){-2.5}{4.0}
\Photon(120,-130)(120,-95){2.0}{5.5}
\Photon(120,-85)(120,-50){2.0}{3.5}
\CArc(120,-90)(5,-90,90)
\CArc(120,-60)(5,0,180)
\Vertex(120,-130){1.6}
\Vertex(120,-50){1.6}
\Vertex(80,-70){1.6}
\Vertex(100,-60){1.6}
\ArrowLine(260,-70)(290,-55)
\ArrowLine(290,-55)(340,-30)
\ArrowLine(300,-90)(260,-70)
\ArrowLine(340,-110)(300,-90)
\ArrowLine(240,-150)(290,-125)
\ArrowLine(290,-125)(340,-150)
\Gluon(210,-70)(260,-70){2.5}{6}
\Gluon(310,-95)(345,-95){2.5}{4.5}
\Photon(290,-125)(290,-90){2.0}{4.5}
\Photon(290,-80)(290,-55){2.0}{3.5}
\CArc(290,-85)(5,-90,90)
\Text(187.5,-175)[cc]{Figure 5: NLO gluon-initiated corrections involving}
\Text(187.5,-190)[cc]{only quark-gluon vertices}
\Vertex(290,-125){1.6}
\Vertex(290,-55){1.6}
\Vertex(260,-70){1.6}
\Vertex(310,-95){1.6}
\end{picture}
\end{center}
\newpage
\begin{center} \begin{picture}(350,155)(10,15)
\Gluon(20,130)(70,130){3}{7.5}
\Gluon(200,130)(250,130){3}{7.5}
\Gluon(45,131)(75,151){2}{3.5}
\Gluon(225,131)(255,151){2}{3.5}
\Vertex(45,131){1.6}
\Vertex(225,131){1.6}
\ArrowLine(160,166)(70,130)
\ArrowLine(70,130)(90,122)
\ArrowLine(20,30)(90,58)
\ArrowLine(90,58)(160,30)
\ArrowLine(250,130)(340,94)
\ArrowLine(270,138)(250,130)
\ArrowLine(200,30)(270,58)
\ArrowLine(270,58)(340,30)
\ArrowLine(340,170)(270,138)
\ArrowLine(90,122)(160,94)
\Photon(90,122)(90,58){3}{7.5}
\Photon(270,58)(270,118){3}{7.5}
\Photon(270,126)(270,138){3}{1.5}
\CArc(270,122)(4,-90,90)
\Vertex(70,130){1.6}
\Vertex(92,121.5){1.6}
\Vertex(90,58){1.6}
\Vertex(268,137.5){1.6}
\Vertex(270,58){1.6}
\Vertex(250,130){1.6}
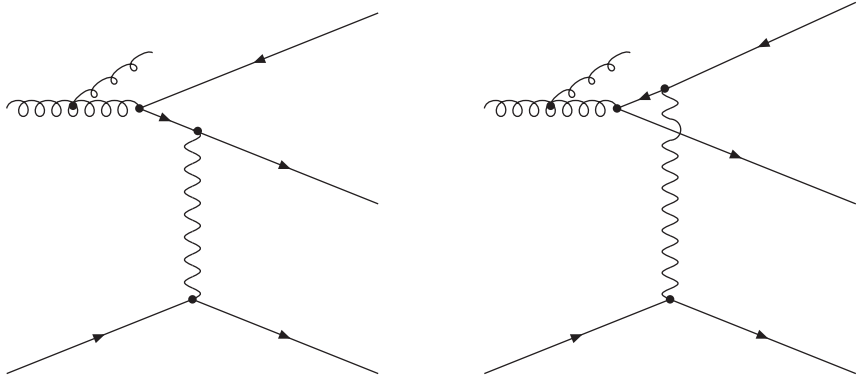
\Text(181,-2)[cc]{Figure 6: NLO gluon-initiated corrections containing}
\Text(181,-16)[cc]{the three gluon vertex}
\end{picture} \end{center}
\vspace{0.69 in}
The contribution of those diagrams to the electron-proton \( \frac{d^{2}
\sigma}{dx dQ^{2}} \) cross section can be written, in the free parton model,
as a convolution over the free gluon density:
\[ \frac{d^{2} \sigma^{\left( h \bar{h} \right)}_{g}}{dx dQ^{2}} = 
\int_{x}^{1} \frac{d \xi}{\xi} \rho_{g} \left( \xi \right) \frac{d^{2} 
\hat{\sigma}^{\left( h \bar{h} \right)}_{g}}{dz dQ^{2}} \]
where \( \frac{d^{2} \hat{\sigma}_{g}^{\left( h \bar{h} \right)}}{dz dQ^{2}} \)
is the corresponding lepton-gluon cross section. \\
The leading term in the gluon-gluon collinear limit comes from the last 2
diagrams in Figure 6; furthermore it can be written in the same form as the
Born contribution if one replaces the free gluon density \( \rho_{g} \left(
\xi \right) \) in eq.(9) with the next-to-leading order correction
\[ \delta \rho^{(2)}_{g} \left( \xi , \frac{Q^{2}}{Q^{2}_{0}} \right) =
\frac{\alpha_{s}}{2 \pi} \log \left(\frac{Q^{2}}{Q^{2}_{0}} \right) \int_{\xi}^
{1} \frac{d \lambda}{\lambda} P_{gg} \left( \frac{\xi}{\lambda} \right)
\rho_{g} \left( \lambda \right) \]
where the explicit expression for the splitting function \( P_{gg} \left( z
\right) \), valid for \( z < 1 \), is given by \cite{phd}
\begin{equation}
P_{gg} \left( z \right) = 2 C_{2} \left( A \right) \left[ \frac{z}{1 - z} + z
\left( 1 - z \right) \right] \  , \ C_{2} \left( A \right) = N = 3 
\end{equation}
Notice that eq.(12) differs from the standard expression derived in
references \cite{dglap1} \cite{dglap2}, id est
\begin{equation}
P_{gg} \left( z \right) = 2 C_{2} \left( A \right) \left[ \frac{z}{1 - z} +
z \left( 1 - z \right) + \frac{1 - z}{z} \right]
\end{equation}
Moreover we point out that \( P_{gg} \left( z \right) \) in eq.(12) vanishes
linearly as \( z \to 0 \) and that it is not invariant under \( z
\Leftrightarrow 1 - z \) exchange contrary to the previous expression. \\
In addition both \( P_{gf} \left( z \right) \) in eq.(11) and \( P_{gg}
\left( z \right) \) in eq.(12) can be obtained from the corresponding
Altarelli-Parisi results by neglecting all terms that do not vanish as
\( z \to 0 \). \\
Therefore the NLO gluon-gluon collinear divergences can be regularized and then
reabsorbed in the following redefinition of the gluon density in the ``Born
level'' term (eq.(9))
\begin{equation}
\rho_{g} \left( \xi \right) \Rightarrow \rho_{g} \left( \xi \right) + \delta
\rho_{g}^{(2)} \left( \xi , \frac{Q^{2}}{Q_{0}^{2}} \right)
\end{equation}
Finally \( \delta \rho^{(2)}_{g} \left( \xi , \frac{Q^{2}}{Q^{2}_{0}} \right)
\) turns out to be divergent in view of a soft singularity in the limit \(
\frac{\xi}{\lambda} \to 1 \) (gluon emitted with arbitrarily small energy). \\
It is a peculiar feature of gauge theories due to the fact that gauge bosons
are massless. \\
If one examines the NLO interference terms of the virtual corrections with the
Born contributions for \( e \ g \rightarrow e \ q_{h} \ \bar{q}_{h} \)
transition, related to collinear divergences, soft singularities are found,
which compensate the analogous divergences in the real emission contributions.
\\
This mechanism is similar to the Bloch-Nordsieck compensation well known in
QED \cite{fact1} \cite{blonor}. \\
The mentioned virtual corrections are given by the six diagrams in figures 7
and 8 (three more diagrams are obtained from those in figure 7 by reversing the
heavy quark line). \\
The loops in figure 7 involve only massless states; they are indeed scaleless
integrals which vanish if appropriately defined in dimensional regularization
\cite{fact1}. \\
Only three diagrams are left (figure 8); their interference terms with the
Born contributions have been included in the calculation \cite{phd}. \\ 
In conclusion the overall next-to-leading order correction to \( \frac{d^{2}
\sigma}{dx dQ^{2}} \), for the heavy flavour \( h \) production (via
boson-gluon-fusion), computed in the collinear limit, can be reabsorbed in
the Born contribution by means of the following redefinition of the free gluon
density in eq.(9):
\begin{equation}
\rho_{g} \left( \xi \right) \Rightarrow \rho_{g} \left( \xi \right) +
\delta \rho_{g} \left( \xi , \frac{Q^{2}}{Q^{2}_{0}} \right)
\end{equation}
with
\normalsize
\begin{equation}
\delta \rho_{g} \left( x , \frac{Q^{2}}{Q^{2}_{0}} \right) = \frac{\alpha_{s}
}{2 \pi} \log \left( \frac{Q^{2}}{Q_{0}^{2}} \right) \int_{x}^{1}
\frac{d \xi}{\xi} \left[ \sum_{f} P_{gf} \left( \frac{x}{\xi} \right)
\rho_{f} \left( \xi \right) +  P_{gg} \left( \frac{x}{\xi} \right) \rho_{g}
\left( \xi \right) \right] 
\end{equation}
\large
The splitting functions \( P_{gf} \left( z \right) \) and \( P_{gg}  \left(
z \right) \) are given by
\[ P_{gf} \left( z \right) = C_{2} \left( F \right) z \ , \ P_{gg} \left( z
\right) = 2 C_{2} \left( A \right) z \left[ \frac{1}{\left( 1 - z \right)_{+}}
+ \left( 1 - z \right) \right] \]
The sum is understood to be over all light (anti)quarks. \\
\begin{center}   \begin{picture}(81,50)(120,140)
\ArrowLine(140,185)(70,150)
\ArrowLine(70,150)(90,140)
\ArrowLine(90,140)(140,115)
\ArrowLine(40,65)(90,90)
\ArrowLine(90,90)(140,65)
\ArrowLine(310,185)(240,150)
\ArrowLine(240,150)(260,140)
\ArrowLine(260,140)(310,115)
\ArrowLine(210,65)(260,90)
\ArrowLine(260,90)(310,65)
\ArrowLine(225,45)(155,10)
\ArrowLine(155,10)(175,0)
\ArrowLine(175,0)(225,-25)
\ArrowLine(125,-75)(175,-50)
\ArrowLine(175,-50)(225,-75)
\Photon(90,90)(90,140){2.0}{5.5}
\Photon(260,90)(260,140){2.0}{5.5}
\Photon(175,-50)(175,0){2.0}{5.5}
\Gluon(0,150)(25,150){2.0}{2.5}
\Gluon(45,150)(70,150){2.0}{2.5}
\Gluon(170,150)(195,150){2.0}{2.5}
\Gluon(215,150)(240,150){2.0}{2.5}
\Gluon(85,10)(110,10){2.0}{2.5}
\Gluon(130,10)(155,10){2.0}{2.5}
\GlueArc(35,150)(10,0,180){2.0}{3.14}
\GlueArc(35,150)(10,180,360){2.0}{3.14}
\DashCArc(120,10)(10,5,175){3.14}
\DashCArc(120,10)(10,185,355){3.14}
\CArc(205,150)(10,0,360)
\Vertex(25,150){1.6}
\Vertex(45,150){1.6}
\Vertex(70,150){1.6}
\Vertex(90,140){1.6}
\Vertex(195,150){1.6}
\Vertex(215,150){1.6}
\Vertex(240,150){1.6}
\Vertex(260,140){1.6}
\Vertex(110,10){1.6}
\Vertex(130,10){1.6}
\Vertex(155,10){1.6}
\Vertex(175,0){1.6}
\Vertex(90,90){1.6}
\Vertex(260,90){1.6}
\Vertex(175,-50){1.6}
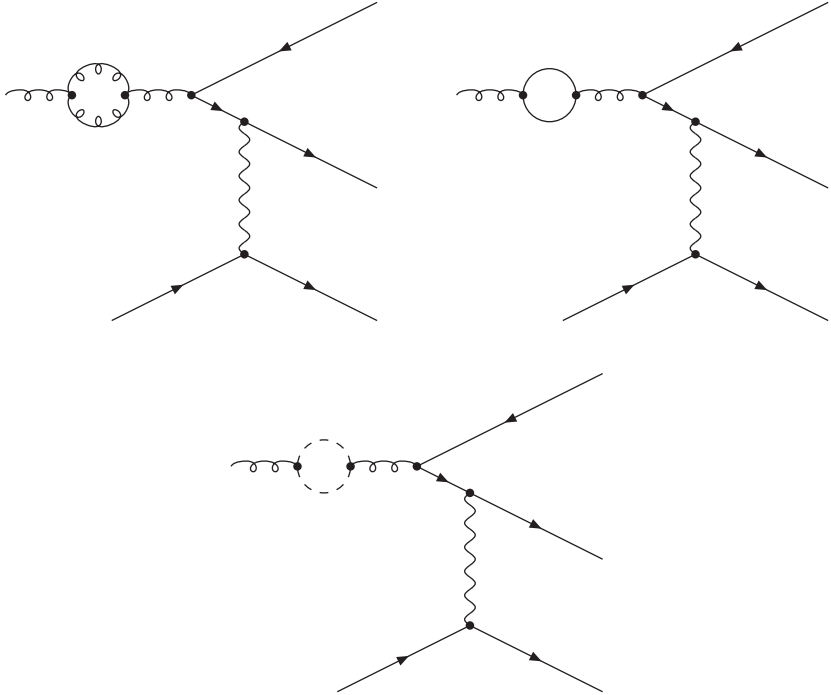
\Text(155,-105)[cc]{Figure 7: Gluon self energy contributions of order
\( \alpha_{s} \)}
\end{picture}   \end{center}
\vspace{3.65 in}
Therefore the effective gluon density involved in heavy flavours
production in e-p DIS, at intermediate values of the factorization scale,
follows a different evolution equation with respect to usual DGLAP eqs.;
needless to say this fact affects the (anti)quarks densities too as soon as the
evolution equations are coupled. \\
This feature is due to the assumption on the dominance of boson-gluon-fusion
mechanism; it is indeed a consequence of the explicit flavour symmetry breaking
(intrinsic heavy flavours contributions are suppressed) and it is expected to
be verified at moderate values of the scale \( \mu \) with respect to
\( m_{h} \). \\
For \( \mu \) values much higher mass effects are negligible and the heavy
flavour \( h \) behaves as a light (massless) flavour; the intrinsic
contributions to the cross sections can no longer be neglected. In that case
the gluon density follows again the standard DGLAP evolution equation so as
the density for \( h \) quarks. \\
\begin{center}   \begin{picture}(-312,-245)(310,105)
\ArrowLine(70,80)(130,120)
\ArrowLine(90,26.666)(70,40)
\ArrowLine(130,0)(90,26.666)
\ArrowLine(70,40)(70,80)
\ArrowLine(40,-48.333)(90,-15)
\ArrowLine(90,-15)(140,-48.333)
\ArrowLine(235,80)(255,93.333)
\ArrowLine(255,93.333)(295,120)
\ArrowLine(295,0)(235,40)
\ArrowLine(235,40)(235,80)
\ArrowLine(205,-48.333)(255,-15)
\ArrowLine(255,-15)(305,-48.333)
\ArrowLine(175,-153.333)(165,-126.666)
\ArrowLine(215,-180)(175,-153.333)
\ArrowLine(155,-100)(215,-60)
\ArrowLine(165,-126.666)(155,-100)
\ArrowLine(115,-228.333)(165,-195)
\ArrowLine(165,-195)(215,-228.333)
\Photon(90,-15)(90,26.666){2.0}{4.5}
\Photon(255,-15)(255,21.666){2.0}{3.5}
\Photon(255,31.666)(255,93.333){2.0}{6.5}
\Photon(165,-195)(165,-151.666){2.0}{5.0}
\Photon(165,-141.666)(165,-126.666){-2.0}{2.0}
\Gluon(0,60)(40,60){2.0}{4.5}
\Gluon(40,60)(70,80){2.0}{4.0}
\Gluon(40,60)(70,40){2.0}{4.0}
\Gluon(165,60)(205,60){2.0}{4.5}
\Gluon(205,60)(235,80){2.0}{4.0}
\Gluon(205,60)(235,40){2.0}{4.0}
\Gluon(85,-120)(125,-120){2.0}{4.5}
\Gluon(125,-120)(155,-100){2.0}{4.0}
\Gluon(125,-120)(175,-153.333){2.0}{6.5}
\CArc(255,26.666)(5,-90,90)
\CArc(165,-146.666)(5.0,90,-90)
\Vertex(40,60){1.6}
\Vertex(70,40){1.6}
\Vertex(70,80){1.6}
\Vertex(90,-15){1.6}
\Vertex(90,26.666){1.6}
\Vertex(255,-15){1.6}
\Vertex(255,93.333){1.6}
\Vertex(205,60){1.6}
\Vertex(235,80){1.6}
\Vertex(235,40){1.6}
\Vertex(165,-195){1.6}
\Vertex(165,-126.666){1.6}
\Vertex(125,-120){1.6}
\Vertex(175,-153.333){1.6}
\Vertex(155,-100){1.6}
\Text(152,-257)[cc]{Figure 8: \( e \ g \rightarrow e \ q_{h} \ \bar{q}_{h} \)
virtual corrections}
\end{picture}  \end{center}

\end{document}